\documentclass[useAMS]{mn2e}
 \usepackage{times}
 \usepackage{epsfig}
 \usepackage{amssymb}

 \title[Plutino 15810 (1994 JR$_1$), a quasi-satellite of Pluto]
       {Plutino 15810 (1994 JR$_1$), an accidental quasi-satellite of Pluto}

 \author[C. de la Fuente Marcos and R. de la Fuente Marcos]
        {C.~de~la~Fuente~Marcos\thanks{E-mail: nbplanet@fis.ucm.es}
         and
         R. de la Fuente Marcos \\
         Universidad Complutense de Madrid,
         Ciudad Universitaria, E-28040 Madrid, Spain}
 \date{Accepted 2012 September 10.
       Received 2012 August 21;
       in original form 2012 July 29}
 \pubyear{XXXX}
 
\begin{document}
  \maketitle

  \begin{abstract}
     In the solar system, quasi-satellites move in a 1:1 mean motion resonance going around their 
     host body like a retrograde satellite but their mutual separation is well beyond the Hill 
     radius and the trajectory is not closed as they orbit the Sun not the host body. Although they 
     share the semi-major axis and the mean longitude of their host body, their eccentricity and 
     inclination may be very different. So far, minor bodies temporarily trapped in the 
     quasi-satellite dynamical state have been identified around Venus, Earth, the dwarf planet (1) 
     Ceres, the large asteroid (4) Vesta, Jupiter, Saturn and Neptune. Using computer simulations, 
     Tiscareno and Malhotra have predicted the existence of a small but significant population of 
     minor bodies moving in a 1:1 mean motion resonance with Pluto. Here we show using $N$-body 
     calculations that the Plutino 15810 (1994 JR$_1$) is currently an accidental quasi-satellite of 
     Pluto and it will remain as such for nearly 350,000 yr. By accidental we mean that the 
     quasi-satellite phase is triggered (or terminated) not by a direct gravitational influence in 
     the form of a discrete close encounter but as a result of a resonance. The relative mean 
     longitude of the Plutino circulates with a superimposed libration resulting from the 
     oscillation of the orbital period induced by the 2:3 mean motion resonance with Neptune. These 
     quasi-satellite episodes are recurrent with a periodicity of nearly 2 Myr. This makes 15810 the 
     first minor body moving in a 1:1 mean motion resonance with Pluto and the first quasi-satellite 
     found in the trans-Neptunian region. It also makes Pluto the second dwarf planet, besides 
     Ceres, to host a quasi-satellite. Our finding confirms that the quasi-satellite resonant phase 
     is not restricted to small bodies orbiting major planets but is possible for dwarf 
     planets/asteroids too. Moreover, 15810 could be considered as a possible secondary target for 
     NASA's Pluto-Kuiper Belt Mission New Horizons after the main Pluto flyby in 2015. This opens 
     the possibility of studying first hand and for the first time a minor body in the 
     quasi-satellite dynamical state.
  \end{abstract}


  \section{Introduction}
     Quasi-satellites are minor bodies that share the semi-major axis and the mean longitude of their host body but may have different 
     eccentricity and inclination. They move like a retrograde satellite but their separation is well outside the Hill sphere of the 
     host body and the trajectory is not closed; therefore, they are not bound satellites as they orbit the Sun not the host body. The 
     theory behind these remarkable objects was first studied in 1913 (Jackson 1913) but the topic was largely neglected by the 
     scientific community until the end of the XXth century when co-orbital motion of planets and asteroids started receiving more 
     attention on theoretical grounds (Mikkola \& Innanen 1997; Wiegert, Innanen \& Mikkola 2000). So far, such quasi-satellites have 
     been found around Venus (Mikkola et al. 2004), Earth (Wiegert, Innanen \& Mikkola 1997; Connors et al. 2002; Connors et al. 2004; 
     Brasser et al. 2004; Christou \& Asher 2011), the dwarf planet (1) Ceres and the large asteroid (4) Vesta (Christou \& Wiegert 
     2012), Jupiter (Kinoshita \& Nakai 2007; Wajer \& Kr\'olikowska 2012), Saturn (Gallardo 2006) and Neptune (de la Fuente Marcos \&
     de la Fuente Marcos 2012a).  
     \hfil\par
     If quasi-satellites have been discovered orbiting rocky bodies in the inner solar system and theory predicts long-term stability 
     for quasi-satellite orbits in the outer solar system, one may wonder whether a quasi-satellite could orbit an object like the 
     dwarf planet Pluto. The existence of a small but significant population of minor bodies experiencing co-orbital resonant behaviour 
     with respect to Pluto in the form of libration or slow circulation of the relative mean longitude has been predicted in the 
     context of chaotic diffusion of trans-Neptunian objects (Yu \& Tremaine 1999; Tiscareno \& Malhotra 2009). Such objects may 
     experience relatively close, low velocity encounters with Pluto which translate into comparatively large perturbational effects if
     they occupy the Kozai resonance (Tiscareno \& Malhotra 2009). Using data from the JPL's HORIZONS system\footnote{http://ssd.jpl.nasa.gov/?horizons} 
     we performed a numerical survey looking for minor bodies relatively close to Pluto during the next few years in order to find 
     candidates for possible co-orbital resonant behaviour. Our search resulted in a promising candidate, 15810 (1994 JR$_1$). We 
     identified 15810 as currently located at about 3.1 AU from Pluto and slowly approaching to its peripluto at 2.7 AU within a 
     timeframe of 5 years. The object was originally discovered with the 2.5 m Isaac Newton Telescope at La Palma on May 12, 1994 
     (Irwin et al. 1994; Irwin, Tremaine \& \.{Z}ytkow 1995) and it has a diameter of 251 km (Irwin et al. 1995). Its orbit, which is 
     quite reliable, has been computed using 43 observations with an arc-length of 2236 
     days\footnote{http://www.minorplanetcenter.net/db\_search/show\_object?object\_id= 1994+JR1} 
     and its $UBVRI$ colors have also been obtained (Barucci et al. 1999).
     \hfil\par
     In this Letter and with the help of $N$-body calculations, we show that the Plutino asteroid 15810 (1994 JR$_1$) currently follows 
     a quasi-satellite path around Pluto. This Letter is organized as follows: in Section 2, we briefly outline our numerical model. 
     Section 3 presents and discusses our results. Our conclusions are summarized in Section 4.

  \section{Numerical model}
     In order to investigate the orbital evolution of 15810 and its possible co-orbital resonant behaviour with respect to Pluto we 
     have performed $N$-body calculations in both directions of time. The numerical integrations of the orbit of Plutino 15810 
     presented here were computed with the Hermite integrator (Makino 1991; Aarseth 2003), in a model solar system which included the 
     perturbations by the eight major planets (Mercury to Neptune) and treat the Earth and the Moon as two separate objects, it also 
     includes the barycentre of the dwarf planet Pluto-Charon system and the three largest asteroids, (1) Ceres, (2) Pallas and (4) 
     Vesta. The standard version of this direct $N$-body code is publicly available from the IoA web 
     site\footnote{http://www.ast.cam.ac.uk/$\sim$sverre/web/pages/nbody.htm}; additional details can be found in de la Fuente Marcos
     \& de la Fuente Marcos (2012b).
     Results in the figures have been obtained using initial conditions (positions and velocities in the barycentre of the 
     solar system referred to the JD2456200.5 epoch) provided by the JPL Horizons system (Giorgini et al. 1996; Standish 1998). In 
     addition to the calculations completed using the nominal orbital elements in Table \ref{orb} we have performed 100 control 
     simulations using sets of orbital elements obtained from the nominal ones using the accepted uncertainties (3$\sigma$). The 
     sources of the Heliocentric Keplerian osculating orbital elements of 15810 are the JPL Small-Body 
     Database\footnote{http://ssd.jpl.nasa.gov/sbdb.cgi} and the AstDyS-2 portal\footnote{http://hamilton.dm.unipi.it/astdys/}.
     
%
%
%
         \begin{table}
            \fontsize{8} {11pt}\selectfont
            \tabcolsep 0.10truecm
            \caption{Heliocentric Keplerian orbital elements of 15810 (1994 JR$_1$) used in this research. 
                     Values include the 1-$\sigma$ uncertainty.
                     (Epoch = JD2456200.5, 2012-Sep-30.0; J2000.0 ecliptic and equinox.
                     Source: JPL Small-Body Database and AstDyS-2 portal.)} 
            \begin{tabular}{ccc}
              \hline
              Semi-major axis, $a$                      & = & 39.24$\pm$0.02 AU \\
              Eccentricity,   $e$                       & = & 0.1143$\pm$0.0003 \\
              Inclination,    $i$                       & = & 3.8032$\pm$0.0002 $^{\circ}$ \\
              Longitude of ascending node, $\Omega$     & = & 144.753$\pm$0.011 $^{\circ}$ \\
              Argument of perihelion, $\omega$          & = & 102.1$\pm$0.2 $^{\circ}$ \\
              Mean anomaly, $M$                         & = & 24.42$\pm$0.12 $^{\circ}$ \\ 
              \hline
            \end{tabular}
            \label{orb}
         \end{table}
%
%

  \section{Orbital evolution}
\begin{figure}
       \centering
        \includegraphics[width=\linewidth]{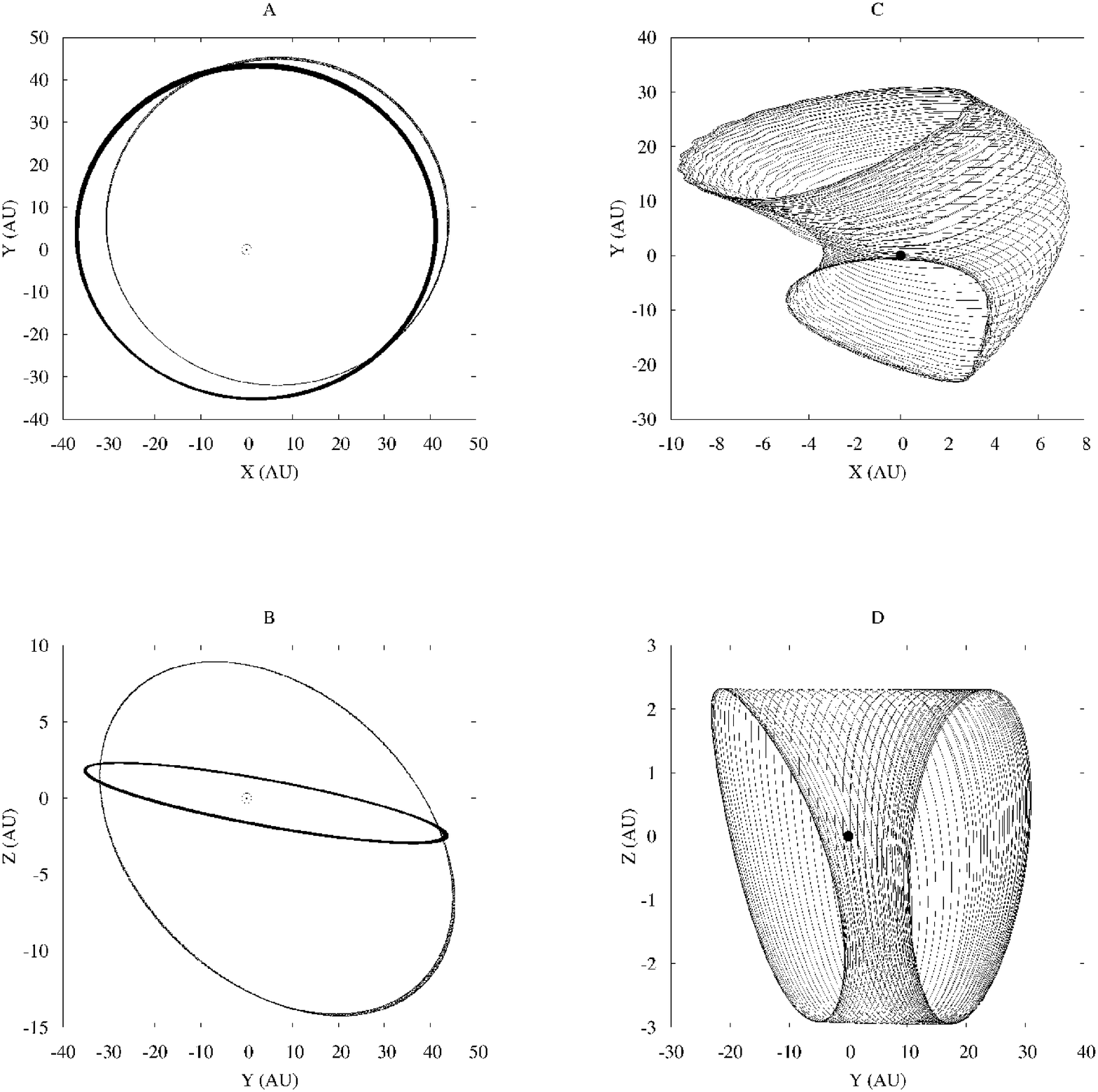}
        \caption{Orbital evolution of Plutino 15810 (1994 JR$_1$) in the time interval (0, 15,000) yr. {\bf A-B}. 
                 The orbits of 15810 (thick line) and Pluto (the barycentre of the Pluto-Charon system) when seen 
                 in the heliocentric J2000 ecliptic frame of reference with the x-axis aligned towards the vernal 
                 equinox. In {\bf A} they are seen projected from the direction of the north ecliptic pole 
                 (ecliptic plane). In {\bf B} the orbits are seen projected from the direction of the vernal 
                 equinox. The relatively large difference between the orbital inclination of 15810 
                 (3.80$^{\circ}$) and Pluto (17.14$^{\circ}$) is evident in the figure. {\bf C-D}. The orbit of 
                 15810 in plutocentric coordinates co-rotating with Pluto (the barycentre of the Pluto-Charon 
                 system). The path oscillates in such a way that when the relative mean longitude librates around 
                 0$^{\circ}$ (see Fig. \ref{f2}), Pluto remains inside the path of 15810. The orientation 
                 of the asteroid's orbit allows the path to overlap the position of Pluto without any danger of 
                 collision although close approaches are certainly possible (see the text). In all the figures
                 the origin of time = JD2456200.5, 2012-Sep-30.0. 
                }
        \label{f1}
     \end{figure}

\begin{figure}
       \centering
        \includegraphics[width=\linewidth]{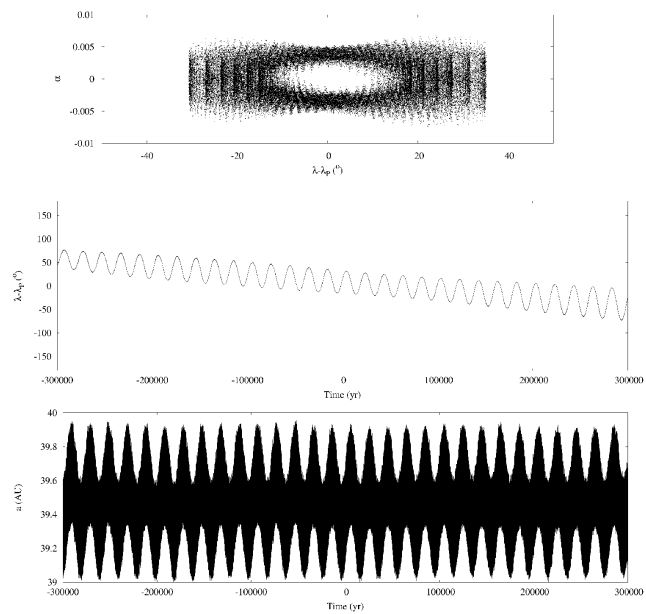}
        \caption{Resonant evolution of Plutino 15810 (1994 JR$_1$). (top) The relative deviation of its 
                 semi-major axis from that of Pluto (the barycentre of the Pluto-Charon system), $\alpha$, 
                 as a function of the difference between its mean longitude from that of Pluto (the 
                 barycentre of the Pluto-Charon system) during the time interval (-25,000, 100,000). The 
                 relative mean longitude librates around 0$^{\circ}$ which is the signpost of the 
                 quasi-satellite behaviour (Mikkola et al. 2006). (middle) Mean longitude relative to 
                 Pluto, $\lambda-\lambda_P$, over the time interval (-300,000, 300,000). Plutino 15810 
                 is currently following a quasi-satellite path with the relative mean longitude librating
                 asymmetrically around 0$^{\circ}$. (bottom) Semi-major axis evolution. The $\sim$20,000 yr 
                 periodic variations induced by the 2:3 mean motion resonance with Neptune are observed in 
                 the evolution of the semi-major axis. Initial conditions (nominal orbit) 
                 are given in Table \ref{orb}.
                }
        \label{f2}
     \end{figure}

     The heliocentric orbits of both 15810 and the barycentre of the Pluto-Charon system (panels {\bf A} and {\bf B}) as well as the 
     quasi-satellite motion features (panels {\bf C} and {\bf D}) are shown in Fig. \ref{f1}. The motion of 15810 from 2012 to 17,012 
     (the origin of time = JD2456200.5, 2012-Sep-30.0) shows quasi-satellite loops ("corkscrew" orbits) as viewed from above Pluto 
     ({\bf C}) and from a point outside Pluto's orbit looking past Pluto in towards the Sun ({\bf D}), in a frame of reference revolving 
     with Pluto. Each loop takes one Plutonian year, 247.7 yr. In order to further study the resonant properties of the path shown in 
     Fig. \ref{f1}, {\bf C-D} panels, let us define the relative deviation of the semi-major axis from that of Pluto by 
     $\alpha = (a - a_P) / a_P$, where $a$ and $a_P$ are the semi-major axes of 15810 and Pluto respectively; and also the relative
     mean longitude $\lambda - \lambda_P$, where $\lambda$ and $\lambda_P$ are the mean longitudes of 15810 and Pluto respectively. In 
     the top panel, Fig. \ref{f2}, the evolution of $\alpha$ as a function of $\lambda - \lambda_P$ during the time interval (-25,000, 
     100,000) yr is displayed. The short period fluctuations are associated to the period of Pluto. In principle, the secular motion is 
     a quasi-harmonic oscillation of the variables $\lambda - \lambda_P$ and $\alpha$; this is the main feature of the quasi-satellite 
     motion (Mikkola et al. 2006). In the middle panel, the mean longitude of 15810 relative to Pluto is displayed. The object 
     currently librates asymmetrically around 0$^{\circ}$ with amplitude 40$^{\circ}$-50$^{\circ}$ and a period of about 20,000 yr that 
     coincides with the libration period of the main resonant angle of a 2:3 mean motion resonance with Neptune (see below). Here by 
     amplitude we mean the difference between the maximum and the minimum values of the relative mean longitude in a period. The object 
     slowly drifted from the neighbourhood of L$_4$ into the quasi-satellite dynamical state and, in the future, it will move towards 
     L$_5$. L$_4$ is the Lagrange point 60$^{\circ}$ ahead of Pluto, L$_5$ is the Lagrange point trailing Pluto 60$^{\circ}$; in 
     general, motion around the Lagrange triangular points follows a tadpole orbit and the objects in these orbits are called Trojans. 
     \hfil\par
\begin{figure}
       \centering
        \includegraphics[width=\linewidth]{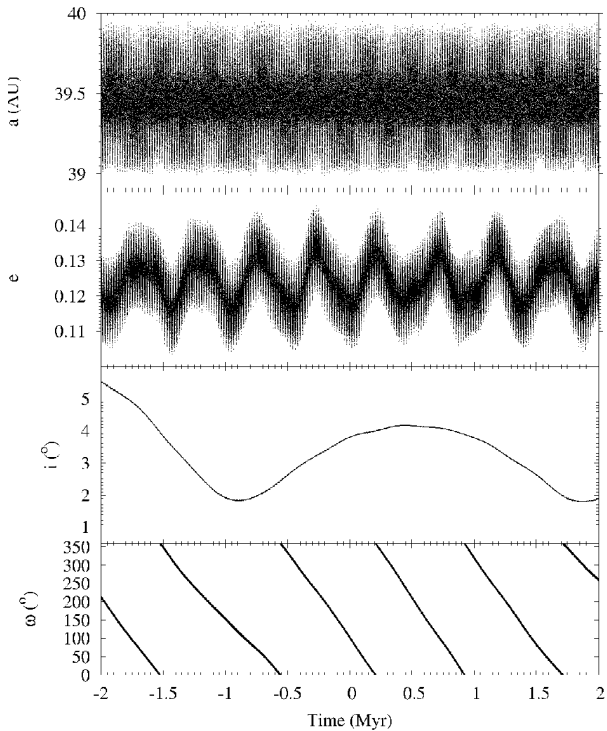}
        \caption{Time evolution of various orbital elements. The orbital elements $a$ (top panel), $e$ (second to top), 
                 $i$ (third to top) and the argument of perihelion (bottom panel) are displayed. Plutino 15810 (1994 
                 JR$_1$)'s orbital period oscillates about a mean value which is exactly 3/2 that of Neptune but its
                 argument of perihelion circulates, it does not librate about a value of 90$^{\circ}$ like in the case
                 of Pluto. Therefore, 15810 is not submitted to a Kozai's secular resonance (Kozai 1962). 
                }
        \label{fele}
     \end{figure}

     Our object, however, does not follow a classical tadpole or horseshoe behaviour after leaving its quasi-satellite path due to the 
     slow circulation likely induced by the close approaches with Pluto. The actual quasi-satellite phase lasts nearly 350,000 yr.  
     Before and after the actual quasi-satellite state, the object does not follow the classical compound orbits resonating between the 
     Trojan and quasi-satellite dynamical states typical of other quasi-satellites (e.g. 2002 VE$_{68}$, Mikkola et al. 2004). These 
     compound orbits have been described on theoretical grounds (Namouni 1999; Namouni, Christou \& Murray 1999). The evolution of the 
     semi-major axis over the plotted periods (Figs. \ref{f2} and \ref{fele}) remains fairly stable and its average value (proper 
     semi-major axis), 39.4518 AU, is very close to that of Pluto, 39.4477 AU. All our simulations suggest that the object has remained 
     in the quasi-satellite phase for nearly 100,000 yr. In the future, 15810 will leave the quasi-satellite path slowly drifting 
     towards the L$_5$ Lagrangian point. This will happen about 250,000 yr from now. The long-term evolution of the orbital elements of 
     15810 displayed in Fig. \ref{fele} shows that the eccentricity exhibits $\sim$0.5 Myr periodic variations, likely the result of an 
     unidentified secular resonance. The inclination also oscillates but the argument of perihelion does not librate about a value of 
     90$^{\circ}$ like in the case of Pluto, it circulates with a period of $\sim$0.7 Myr. Therefore, 15810 is not submitted to a 
     Kozai's secular resonance (Kozai 1962). Close encounters with Pluto have a periodicity of $\sim$2 Myr. The difference between 
     15810's and Pluto's mean longitudes also circulates every $\sim$2 Myr. The similar periodicity strongly suggests that the close 
     approaches with Pluto are responsible for the circulation of the relative mean longitude. If we repeat the calculations assuming
     a negligible mass for the Pluto-Charon system, the librating component of the relative mean longitude (Fig. \ref{f2}, middle 
     panel) vanishes and just the circulating behaviour (but with longer period) remains: Pluto is clearly responsible for the observed 
     quasi-satellite behaviour.
     \hfil\par
%
%
%
%
%
%
%
%
%
%
\begin{figure}
       \centering
        \includegraphics[width=\linewidth]{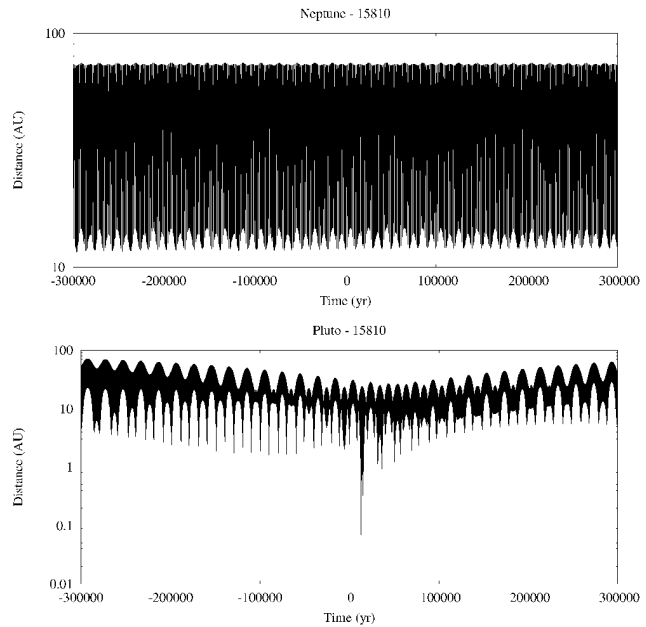}
        \caption{The distance of Plutino 15810 from Neptune (top panel) and from Pluto (the barycentre 
                 of the Pluto-Charon system, bottom panel). The distance of 15810 from Neptune remains 
                 larger than 11.5 AU as it moves in a 2:3 mean motion resonance with Neptune but the 
                 object experiences repeated encounters with Pluto (its barycentre). The Hill radius
                 of Pluto, 0.0385 AU, is also indicated. 
                }
        \label{f3}
     \end{figure}

%
%
     The distance of 15810 from Neptune and Pluto is shown in Fig. \ref{f3}. The distance of 15810 from Neptune remains larger than 
     11.5 AU (the distance from Uranus $>$ 13.5 AU). This indicates that encounters with Neptune (or Uranus) do not cause the object to 
     depart from the quasi-satellite orbit in either direction of time. This is to be expected as 15810 is itself a Plutino, the object 
     is in a 2:3 mean motion resonance with Neptune: in the time period 15810 completes two orbits, Neptune goes around the Sun three 
     times. The resonance argument $\phi = 3 \lambda - 2 \lambda_N - \varpi$, with $\varpi = \Omega + \omega$ and $\lambda_N$ the mean 
     longitude of Neptune, librates about 180$^{\circ}$ with an amplitude of 90$^{\circ}$ (not 82$^{\circ}$ like in the case of Pluto), 
     the libration period is $\sim$20,000 yr. This resonance prevents close encounters between 15810 and Neptune and also between Pluto 
     and Neptune. The slow drift in relative mean longitude observed in Fig. \ref{f2} is not induced by Neptune but by Pluto; the 
     object experiences repeated relatively close encounters with Pluto (more properly, the barycentre of the Pluto-Charon system as in 
     all this discussion) as seen in Fig. \ref{f3}, although initially the closest distances are well outside the Hill sphere of Pluto 
     that has a radius of 0.0385 AU. Approximately 125,000 yr into the past, 15810 had a number of periodic (10,000 yr) periplutos in 
     the range 1.6-1.8 AU. But in the future, in about 13,000 yr, much closer approaches will be possible at 0.07 AU that is less than 
     twice its Hill radius. The object will slowly drift into the area around the L$_5$ Lagrange point.  
     \hfil\par
     The object identified in this Letter is not the classical quasi-satellite found around objects moving in not very eccentric orbits 
     like Venus. The high inclination and eccentricity characteristic of the orbit of Pluto add new complexity to an already 
     challenging dynamical situation. In principle, it may be argued that 15810 is not even co-orbital with Pluto, making the analysis 
     carried out here dynamically unjustified. Following Namouni (1999), the co-orbital region in the case of a host object is defined 
     as $|a - a_O| \leq r_{HO}$, where $r_{HO}$ is the radius of the Hill sphere of the host object and $a_O$, its semi-major axis. In 
     the case of the first bona fide quasi-satellite, 2002 VE$_{68}$, and for the JD2456200.5 epoch, the semi-major axes of the 
     quasi-satellite and Venus are 0.7237 AU and 0.7233 AU respectively, and the Hill radius of Venus is 0.0067 AU. Therefore, 
     Namouni's criterion obviously indicates that 2002 VE$_{68}$ is co-orbital with Venus. The same naive application of the criterion
     to 15810 gives $|39.24 - 39.36| > 0.0385 AU$ in violation of the co-orbitality criterion. Therefore and using this approach, 15810 
     is not even co-orbital with Pluto and the quasi-satellite motion pointed out above is not more than a happy coincidence. However, 
     if we use the proper orbital elements instead of osculating Keplerian orbital elements at a particular epoch, the criterion 
     becomes $|39.4518 - 39.4477| < 0.0401 AU$. The proper elements clearly confirm the co-orbital nature (with Pluto) of 15810; 
     Plutino 15810 is truly, albeit somewhat accidental, a quasi-satellite of Pluto. By accidental we mean that the quasi-satellite 
     phase is not triggered or terminated by direct gravitational interaction in the form of a discrete close encounter (with a certain 
     planet like the Earth in the case of 2002 VE$_{68}$) but as a result of a resonance. The relative mean longitude of the Plutino
     circulates with a superimposed libration resulting from the oscillation of the orbital period induced by the 2:3 mean motion 
     resonance with Neptune. These quasi-satellite episodes are recurrent with a periodicity of nearly 2 Myr. The behaviour found in 
     our calculations have been previously described in numerical simulations (Yu \& Tremaine 1999; Tiscareno \& Malhotra 2009) which 
     predict that only 7\% of objects is expected to experience persistent circulation of the relative mean longitude with respect to 
     Pluto; therefore, the identification of one of these unusual objects provides a useful constraint for models studying the dynamics 
     of the outer solar system as well as giant planet migration. Regarding the origin of 15810, its currently very stable orbit 
     suggests that it is not relatively recent debris originated in collisions within Pluto's system but perhaps a primordial Plutino 
     formed around the same epoch Pluto came into existence. The object studied here may be part of an outer solar system analogue to 
     the population of Main Belt asteroids recently found co-orbiting with the dwarf planet (1) Ceres and the large asteroid (4) Vesta 
     (Christou \& Wiegert 2012). Although the figures have been computed using the nominal orbit in Table \ref{orb}, the other 
     simulations gave very similar results, over the time interval shown.  
%
%
%
%

  \section{Conclusions}
     Pluto's system continues being a source of controversy, unanswered questions and surprises more than 80 years after its discovery. 
     Pluto's planethood demotion in August 2006 still stirs debate today and the recent finding of a fifth moon orbiting Pluto by the 
     HST\footnote{http://hubblesite.org/newscenter/archive/releases/solar-system/pluto/2012/32/} just confirms the unexpectedly complex 
     nature of the system. In our work, we show that 15810 currently follows a quasi-satellite orbit relative to Pluto; therefore and 
     besides having 5 regular satellites, Pluto has at least one quasi-satellite. This makes 15810 the first minor body found moving in 
     a 1:1 mean motion resonance with Pluto and the first quasi-satellite found in the trans-Neptunian region of the solar system. It 
     also makes Pluto the second dwarf planet, besides Ceres, to host a quasi-satellite. Our finding also confirms that the 
     quasi-satellite resonant phase is not restricted to small bodies orbiting major planets but it is possible for dwarf 
     planets/asteroids too. We also provide a new and somewhat unexpected mechanism to land minor bodies into the quasi-satellite 
     dynamical state. On the other hand, Plutino 15810 is a natural candidate for a spacecraft rendezvous mission in the framework of 
     NASA's Pluto-Kuiper Belt Mission New Horizons that is going to complete a flyby with Pluto in 2015 and then continue to explore 
     one or more nearby trans-Neptunian objects in the time frame 2016-2020. It is moving in such an orbit that this object could be a 
     good candidate for a body dynamically related to the Pluto-Charon formation event, and the determination of the physical 
     properties of its surface by spectroscopic observations could be interesting in that respect. If 15810 is selected for the 
     extended mission it will open the possibility of studying in detail for the first time a minor body in the quasi-satellite 
     dynamical state.

  \section*{Acknowledgements}
     The authors would like to thank S. Aarseth for providing the codes used in this research. This work was partially 
     supported by the Spanish 'Comunidad de Madrid' under grant CAM S2009/ESP-1496 (Din\'amica Estelar y Sistemas 
     Planetarios). We thank Dr. Mar\'{\i}a Jos\'e Fern\'andez-Figueroa, Dr. Manuel Rego Fern\'andez and the Department 
     of Astrophysics of the Universidad Complutense de Madrid for providing excellent computing facilities. Most of 
     the calculations and part of the data analysis were completed on the 'Servidor Central de C\'alculo' of the 
     Universidad Complutense of Madrid and we thank the computing staff (Santiago Cano Als\'ua) for help during this 
     stage. In preparation of this Letter, we made use of the NASA Astrophysics Data System, the ASTRO-PH e-print 
     server and the Minor Planets Center data server.

\end{document}